\documentclass[aps,reprint,showpacs,groupedaddress,superscriptaddress,footinbib,longbibliography]{revtex4-2}

\usepackage{mathrsfs}
\usepackage{natbib,hyperref}
\setcitestyle{square,sort&compress,comma,numbers}
\hypersetup{colorlinks=true, citecolor=blue, urlcolor=blue, linkcolor=blue}
\usepackage[english]{babel}
\usepackage[utf8]{inputenc}
\usepackage{graphicx}
\usepackage[caption=false]{subfig}
\usepackage{amsmath}
\usepackage{amsfonts}
\usepackage{amssymb}
\usepackage{soul}
\usepackage{cancel}
\setulcolor{red}
\usepackage{easyReview}
\usepackage{xcolor}
\usepackage[normalem]{ulem}
\usepackage{placeins}

\begin{document}
\title{Learning Effective Hydro-Phoretic Interactions in Active Matter}

\author{Palash Bera}
\affiliation{Institute for Condensed Matter Physics, Technische Universit{\"a}t Darmstadt, Hochschulstraße 8, 64289 Darmstadt, Germany.}

\author{Aritra K. Mukhopadhyay}
\affiliation{Institute for Condensed Matter Physics, Technische Universit{\"a}t Darmstadt, Hochschulstraße 8, 64289 Darmstadt, Germany.}

\author{Benno Liebchen}
\email{benno.liebchen@pkm.tu-darmstadt.de}
\affiliation{Institute for Condensed Matter Physics, Technische Universit{\"a}t Darmstadt, Hochschulstraße 8, 64289 Darmstadt, Germany.}

\date{\today}

\begin{abstract}
In the quest to understand large-scale collective behavior in active matter, the complexity of hydrodynamic and phoretic interactions remains a fundamental challenge. 
To date, most works either focus on minimal models that do not (fully) account for these interactions, or explore relatively small systems. 
The present work develops a generic method that combines high-fidelity simulations with symmetry-preserving descriptors and neural networks to predict hydro-phoretic interactions directly from particle coordinates (effective interactions). 
This method 
enables, for the first time, self-contained particle-only simulations and theories with full hydro-phoretic two-body interactions.
\end{abstract}


\maketitle

\section{Introduction} 
One of the central fascinations of synthetic active matter is its ability to show intriguing self-organized behaviors in systems that are simpler, more controllable, and better reproducible than their biological counterparts. Prime examples include dynamic and chiral clustering at low density~\cite{GinotNCAggregationfragmentationIndividual2018,PalacciSLivingCrystals2013,TheurkauffPRLDynamicClustering2012,Pohl_PRL_Dynamic_Clustering_2014,ButtinoniPRLDynamicalClustering2013,PengSMSelforganizationActive2024,KetzetziSelfreconfiguringColloidal2024,CapriniCPDynamicalClustering2024,BaconnierRMPSelfaligningPolar2025,AubretNPTargetedAssembly2018,AubretSMDiffusiophoreticDesign2018,AubretNCMetamachinesPluripotent2021,MartinetAISRotationControl2023,liebchen2022chiral,gompper20252025,ZhangNPActivePhase2021,YanNMReconfiguringActive2016a}, 
activity-induced structure formation in active-passive mixtures ~\cite{SteimelPNASEmergentUltra2016,SmrekPRLSmallActivity2017,StenhammarPRLActivityInducedPhase2015,MasonNCDynamicalPatterns2025,BonifaceNCClusteringInduces2024,WilliamsNCConfinementinducedAccumulation2022},
and the formation of active molecules that acquire motility from nonreciprocally interacting building blocks~\cite{SotoPRLSelfAssemblyCatalytically2014,niu2018dynamics,AgudoCanalejoPRLActivePhase2019,SchmidtJCPLightcontrolledAssembly2019,WangNCActiveColloidal2020,GrauerNCActiveDroploids2021,fehlinger2023collective,HaraNonreciprocalInteractions2025}. 

\par While significant progress has been made in identifying possible and highly plausible explanations for the mechanisms underlying these and many other phenomena, the state-of-the-art of our understanding of large scale behavior in active matter still largely hinges on phenomenological minimal models. These models often do not account for the full interactions among active particles to explain experimentally observed phenomena from first principles. In particular, many of the above-quoted phenomena, from dynamic clustering ~\cite{PalacciSLivingCrystals2013,TheurkauffPRLDynamicClustering2012,ButtinoniPRLDynamicalClustering2013,ginot2018aggregation} to rotating gears \cite{AubretNPTargetedAssembly2018,AubretSMDiffusiophoreticDesign2018,AubretNCMetamachinesPluripotent2021,MartinetAISRotationControl2023}, and active molecules~\cite{SchmidtJCPLightcontrolledAssembly2019,WangNCActiveColloidal2020, GrauerNCActiveDroploids2021}
are now known to strongly depend on hydrodynamic and phoretic interactions~\cite{LiebchenJPCMInteractionsActive2021a,LiebchenACRSyntheticChemotaxis2018a,Zottl_ARCMP_Modeling_Active_2023,golestanian2019phoreticactivematter, MoranARFMPhoreticSelfPropulsion2017,IllienCSRFuelledMotion2017,MarchettiRMPHydrodynamicsSoft2013a,BechingerRMPActiveParticles2016a,WangJACSOpenQuestions2023,carrascoCharacterizationNonequilibriumInteractions2025, Wang_NC_Spontaneous_Vortex_2023}, which play a key role across active matter setups, from active colloids \cite{LiebchenJPCMInteractionsActive2021a,LiebchenACRSyntheticChemotaxis2018a,Zottl_ARCMP_Modeling_Active_2023,golestanian2019phoreticactivematter, MoranARFMPhoreticSelfPropulsion2017,IllienCSRFuelledMotion2017,MarchettiRMPHydrodynamicsSoft2013a,BechingerRMPActiveParticles2016a,WangJACSOpenQuestions2023,carrascoCharacterizationNonequilibriumInteractions2025, Wang_NC_Spontaneous_Vortex_2023,NishiguchiNJPFlagellarDynamics2018} to active droplets \cite{MaassARCMPSwimmingDroplets2016a,FengASSelfSolidifyingActive2023,ChenPRLEvolvingMotility2025,ZiethenPRLNucleationChemically2023,MichelinARFMSelfPropulsionChemically2023a,KimACRChemicalProgramming2024}, and ion-exchange driven microswimmers \cite{niu2018modular,niu2018dynamics,liebchen2018unraveling,moller2021shaping}. 
Thus, to understand collective behavior beyond minimal models, one would need to fully resolve hydro-phoretic interactions among all particles.  In practice, fully resolved approaches, in which both hydrodynamic and phoretic fields are computed in the entire fluid domain and the relevant particle-field boundary conditions are accurately resolved at the surfaces of all particles, are currently feasible only for $\lesssim 10$ particles~\cite{VarmaPRFModelingChemohydrodynamic2019a, SharanSPairInteraction2023, nasouriExactPhoreticInteraction2020, NasouriJFMExactAxisymmetric2020, SharifiMoodJFMPairInteraction2016, RojasPerezJFMHydrochemicalInteractions2021, MontenegroJohnson_EPJE_Regularised_Singularity_2015} and $\lesssim 10^2$ particles in dilute suspensions \cite{Gou_JFM_Computational_Framework_2025}.
For sufficiently regular geometries, 
moderate system sizes of $\sim 10^2$ particles~\cite{delmotteScalableMethodModel2024a, Kohl_AiCM_Fast_Accurate_2023} (and $\sim 10^4-10^5$ particles in diffusion-dominated regimes~\cite{Singh_JOSS_PyStokes_Phoresis_2020,Singh_JCP_Competing_Chemical_2019})
can be reached based on reduced surface-based descriptions, such as boundary-integral or Galerkin methods, which approximately reconstruct hydrodynamic and phoretic interactions from analytically known Green’s functions or surface modes.
Alternatively, systems with $\sim 10^3-10^4$ particles have been explored based on Lattice-Boltzmann-models ~\cite{Scagliarini_SM_Unravelling_Role_2020, Yang_NC_Shaping_Active_2024} that approximate particle field boundary conditions on a (static) grid, or,
by representing hydro-phoretic fields by effective particles and using multi-particle collision dynamics~\cite{robertson2018synthetic, huang2017chemotactic}. 
In all these approaches, scalability comes at the expense of significant discretization effects and/or a limited near-field resolution of hydro-phoretic interactions.
Analytical theories describing large-scale collective behavior of active matter, in turn, currently account for hydro-phoretic far-field interactions at most; see e.g.  \cite{LiebchenJPCMInteractionsActive2021a,saha2014clusters,illien2017fuelled,liebchen2017phoretic,StarkACRArtificialChemotaxis2018,LiebchenACRSyntheticChemotaxis2018a}.

\begin{figure*}[ht] 
	\includegraphics[width=1.0\linewidth]{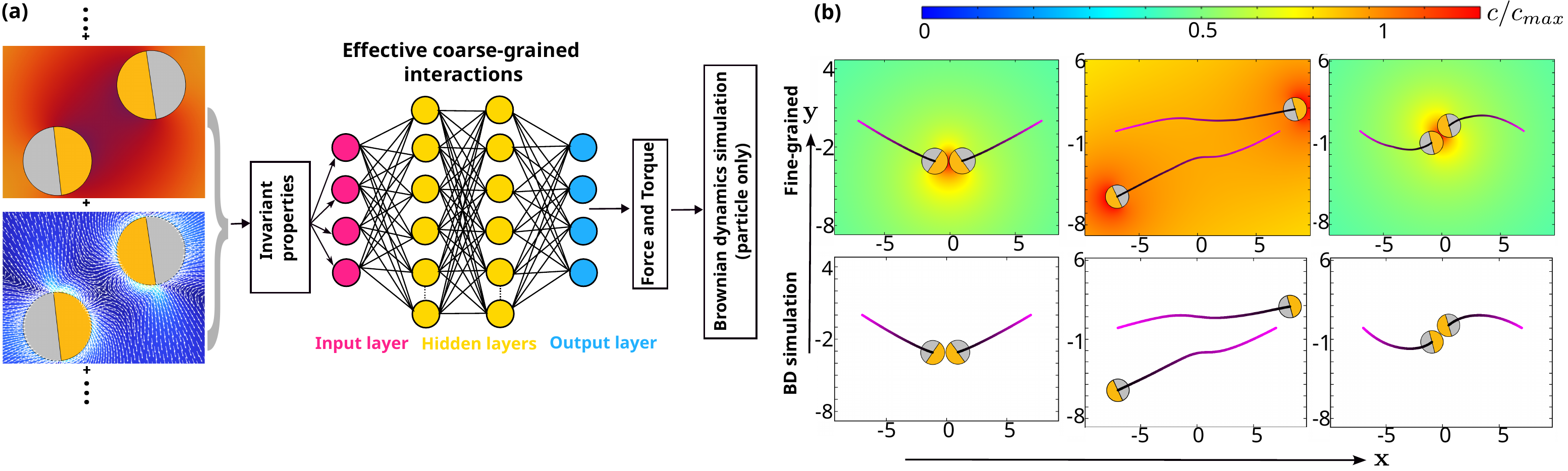}
	\caption{(a) Schematic of the coarse-graining method that extracts symmetry-preserving descriptors from fine-grained simulations (left) and feeds them into a deep artificial neural network (DNN). The DNN is trained to predict effective hydrodynamic and phoretic interactions directly from particle coordinates and enables self-contained Brownian dynamics simulations. (b) Comparison of fine-grained (top) and BD (bottom) trajectories. Background colors show the normalized phoretic field, and trajectory colors indicate time-evolution in arbitrary units. Left and middle panels: $\mu=1$; right panel: $\mu=0$.}
	\label{fig:ml_sch}
\end{figure*}
\par To enable fully realistic analytical theories and very large simulations
of phoretic active matter, we are currently lacking a method to eliminate hydro-phoretic fields from detailed models and to replace their net effect with effective interactions depending on particle coordinates only. 
This situation is in contrast to equilibrium, where effective interactions have been successfully determined for many different problems, based on the systematic reduction of partition functions of multicomponent systems to single species models \cite{likos2001effective}, which are at the heart of much of what we know about equilibrium soft matter physics. In addition, recent machine-learning approaches have advanced the inference of effective interactions in soft and active matter~\cite{RuizGarciaPREDiscoveringDynamic2024,ReesZimmermanSMEffectiveInteractions2025,HemSMLearningGeneral2025,ReesZimmermanJCPNumericalMethods2025,HanPNLearningPhysicsconsistent2022,CichosNMIMachineLearning2020a,terReleJCPMachineLearning2025,HaSRUnravelingHidden2021,JiPRLMachineLearningInteratomic2025,BehlerCRFourGenerations2021,BehlerPRLGeneralizedNeuralNetwork2007,HusicJCPCoarseGraining2020a,WangACSMachineLearning2019a,HanLearningNoisy2025,BayatiInferringSurface2025}. 
In particular, there has been substantial progress in learning the effective interactions of active particles, such as simulated active Brownian particles and experimental Janus colloids~\cite{RuizGarciaPREDiscoveringDynamic2024,ReesZimmermanSMEffectiveInteractions2025,HemSMLearningGeneral2025}. Graph-neural-network approaches have inferred active and two-body forces directly from trajectories of simulated active Brownian particles and experiment with electrophoretic Janus colloids~\cite{RuizGarciaPREDiscoveringDynamic2024}, while inverse-structure methods have obtained activity- and density-dependent effective pair potentials that reproduce the steady-state structure of active Brownian particles~\cite{ReesZimmermanSMEffectiveInteractions2025}. Very recently, stochastic-force-inference methods have learned general pair interactions from experimental trajectories of self-propelled Janus particles and enabled simulations that reproduce experimental observables and extrapolate to different densities~\cite{HemSMLearningGeneral2025}. However, despite these pioneering efforts, the systematic understanding of hydro-phoretic interactions, which govern many remarkable collective phenomena in active matter, remains limited, leaving scalable and quantitatively accurate simulations of large hydro-phoretically interacting active suspensions an open challenge. Currently, it is not yet possible to accurately simulate the large-scale dynamics of hydro-phoretically interacting active systems.

\par In this article, we develop a general method to determine effective hydro-phoretic interactions in active systems, enabling self-contained particle-only models. This framework comprises three parts: (i) high-fidelity simulations resolving near- and far-field hydrodynamic and phoretic couplings, (ii) a machine learning based method to systematically learn an effective, coarse-grained representation of these interactions using invariant descriptors~(Fig. \ref{fig:ml_sch}a), and (iii) a coarse-grained Brownian dynamics (BD) simulation model, where hydro-phoretic interactions are encoded as effective forces and torques~(Fig. \ref{fig:ml_sch}b). This approach massively reduces the complexity of models with hydro-phoretic interactions. It enables a new kind of theories and simulations, treating field-mediated hydro-phoretic couplings as direct interparticle-interactions. 
\section{RESULTS}
\subsection{Main idea}
In this work, we focus on active colloids with instantaneous hydro-phoretic interactions, i.e., low Reynolds number (Stokes regime) and low solute P\'eclet number~\cite{MoranARFMPhoreticSelfPropulsion2017,LiebchenJPCMInteractionsActive2021a,GolestanianNJPDesigningPhoretic2007}, as detailed below. In this regime, the timescale separation between the particle dynamics and the dynamics of the hydro-phoretic fields is complete, such that the instantaneous positions and orientations of all particles fully determine the phoretic and hydrodynamic fields. These fields, in turn, fully determine the velocities and angular velocities of all particles (up to thermal fluctuations and possible additional body forces, e.g., steric interactions). 
Thus, in principle, a mapping exists from particle positions and orientations to velocities and angular velocities (or effective forces and torques), enabling 
self-contained particle-only models. Establishing a method to explicitly determine such a mapping is the main goal of the present work. For simplicity, we focus on two-body interactions in bulk, but note that our approach is generic and can be generalized in the future to account for other factors that can influence hydro-phoretic interactions, such as the presence of substrates, bulk reactions, many-body effects, multiple phoretic fields, and irregular geometries \cite{Michelin_JFM_Phoretic_Selfpropulsion_2014, Das_NC_Boundaries_Can_2015, Simmchen_NC_Topographical_Pathways_2016, Kanso_JCP_Phoretic_Hydrodynamic_2019,Ibrahim_JFM_Multiple_Phoretic_2017}.

We note that simply extracting and tabularizing the interactions from ``all" configurations in the fine-grained simulations would not be feasible: even for a moderately coarse discretization accounting for $100$ interparticle distances and $64$ orientations for each of the three relevant orientation angles, one would need to simulate $\sim 3 \times 10^8$ configurations. In contrast, the present approach allows working directly with a small number of trajectories ($\sim 10^3$). In addition, importantly, and opposed to simple tabularization, the present approach provides a systematic method that can be generalized in the future to account for many-body interactions, substrate, and memory effects. 
\begin{figure*}[ht]
	\includegraphics[width=1.0\linewidth]{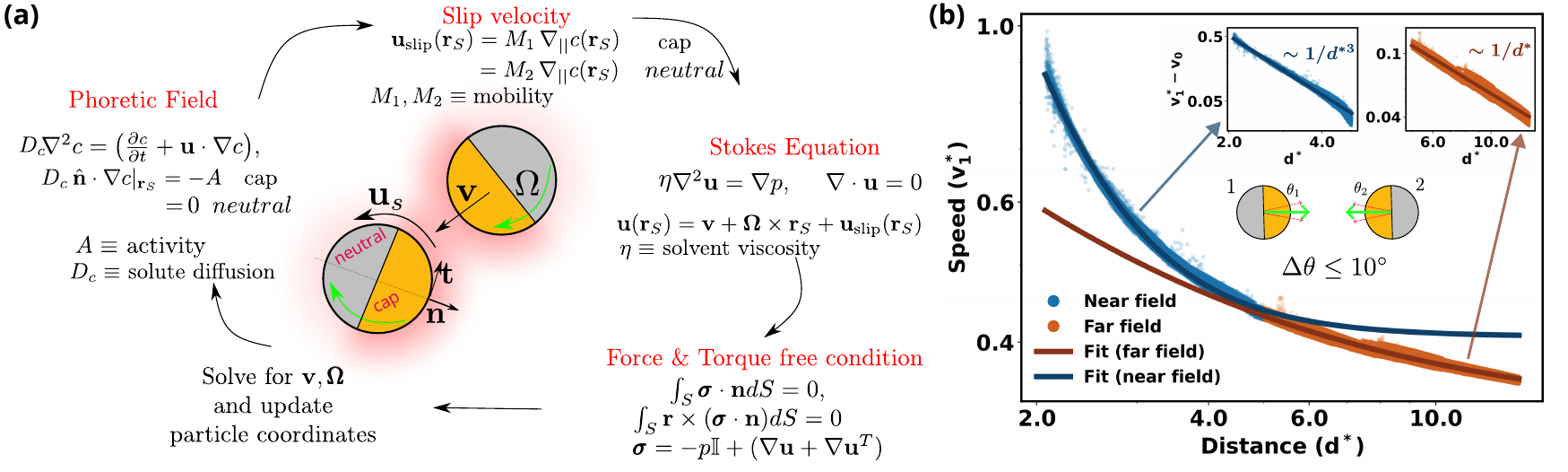}
    \caption{(a) Schematic of the fine-grained model that we solve numerically to determine the full hydrodynamic and phoretic interactions. (b) The dimensionless speed of colloid, $v_1^*$, is fitted to $v_1^{*} = \frac{a}{d^*{^n}} + v_0$, where $d^*$ is the dimensionless separation distance, with fitting parameters $a$, $n$, and $v_0$ (see Appendix for details). The far-field regime is defined for $d>5R$. The parameter $v_0$ corresponds to the self-propulsion speed of an isolated colloid, and the parameter $\mu=1$. The blue and orange data correspond to the near-field and far-field regimes, respectively. The inset plots show the corresponding near- and far-field velocity-decay fits. In the near field, the velocity decays approximately as $\sim 1/d^3$ (fitted exponent $n=3.02$), whereas in the far field the decay is close to $\sim 1/d$, (fitted exponent $n=0.97$). The particle orientations used for this representative case are shown schematically in the inset, with $\Delta\theta\leq10^\circ$}
	\label{fig:sim_sch}
\end{figure*}

\subsection{Model}
We now describe our simulation model, that fully resolves the phoretic and hydrodynamic interactions in two self-diffusiophoretic Janus particles of radius $R$ in two dimensions (2D)
\cite{Michelin_JFM_Phoretic_Selfpropulsion_2014,VarmaPRFModelingChemohydrodynamic2019a, LiebchenJPCMInteractionsActive2021a}
(Fig.~\ref{fig:sim_sch}). 
Each particle, located at position $\mathbf{r}_i$ with orientation $\theta_i$ ($i=\{1,2\}$), features a catalytic hemisphere producing solute at a constant rate $\mathcal{A}(\mathbf{r}_{S_i},t)=A$ and zero elsewhere, where $\mathbf{r}_{S_i}$ denotes points on the particle surface $S_i$. The resulting solute concentration field $c(\mathbf{r},t)$ obeys the advection-diffusion equation $\partial_t c + \mathbf{u}\cdot\nabla c = D_c\nabla^2 c$, where $D_c$ is the solute diffusion coefficient and $\mathbf{u}(\mathbf{r},t)$ the local fluid velocity. We consider the low solute Péclet regime ($Pe_{\text{c}} = R|\mathbf{u}|/D_c \ll 1$), where solute diffusion dominates advection and the phoretic concentration field reaches a quasi-steady state satisfying $D_c\nabla^2 c = 0$, subject to the flux boundary condition $D_c\,\hat{\mathbf{n}}\cdot\nabla c|_{\mathbf{r}_{S_i}} = -\mathcal{A}(\mathbf{r}_{S_i},t)$ and the far-field condition $c\to0$ as $|\mathbf{r}|\to\infty$. Gradients in $c$ generate phoretic slip flows at the particle surfaces, coupling the chemical and hydrodynamic fields. The fluid motion is described by the incompressible Stokes equations $\eta\nabla^2\mathbf{u} = \nabla p$ and $\nabla\cdot\mathbf{u}=0$, where $\eta$, $\mathbf{u}$, and $p$ denote viscosity, velocity, and pressure, respectively. The slip boundary condition $\mathbf{u}_\text{slip} (\mathbf{r}_{S_i}) = \mathcal{M}(\mathbf{r}_{S_i})\,\nabla_{||} c(\mathbf{r}_{S_i},t)$ is applied on the surface of both the particles. This couples the phoretic and hydrodynamic fields, with surface mobility $\mathcal{M}(\mathbf{r}_{S_i})=M_1$ at the catalytic hemisphere and $M_2$ at the neutral hemisphere. In our simulations, we consider both types of particles: with uniform mobility $M_1 = M_2$ and nonuniform mobility: $M_1\neq M_2$. The tangential gradient is denoted by $\nabla_{||}=(\mathbf{I}-\hat{\mathbf{n}}\hat{\mathbf{n}})\cdot\nabla$, where $\hat{\mathbf{n}}$ is the surface normal unit vector. The total surface velocity is $\mathbf{u}(\mathbf{r}_{S_i},t) = \mathbf{v}_i + \mathbf{\Omega}_i\times\mathbf{r}_{S_i} + \mathbf{u}_\text{slip}(\mathbf{r}_{S_i})$, where $\mathbf{v}_i$ and $\mathbf{\Omega}_i$ are the translational and rotational velocities of each particle. Far from the particle, the fluid is quiescent ($\mathbf{u}\to0$ as $|\mathbf{r}|\to\infty$). Because self-propelled particles are force- and torque-free, $\mathbf{v}_i$ and $\mathbf{\Omega}_i$ are determined by solving the coupled hydro-phoretic problem under the constraints of zero net hydrodynamic force and torque, ensuring momentum conservation~\cite{Michelin_JFM_Phoretic_Selfpropulsion_2014,VarmaPRFModelingChemohydrodynamic2019a} (Fig.~\ref{fig:sim_sch}a). 
In the present case, the resulting interactions between the colloids are long-ranged and attractive. 

We choose units of length, mass, time, and concentration as $l_u=R$, $m_u=\eta R^2 D_c /(AM_1)$, $t_u=RD_c /(AM_1)$, and $c_u=AR/D_c$, respectively~\cite{LiebchenJPCMInteractionsActive2021a}. 
This reduces the parameter space to two dimensions, spanned by the nondimensional mobility ratio $\mu=M_2/M_1$, and the solute Péclet number $Pe_c=AM_1R/D_c^2$, which we set to 0, to focus on the regime where phoretic interactions are instantaneous.
We use finite element simulations with an adaptive moving mesh and an implicit backwards differentiation time-stepping formula (see Appendix for details).
To show that our approach does not require a large training dataset, we simulate only $\sim 10^3$ distinct two-particle configurations starting from different relative orientations at fixed separations. We record the instantaneous positions and velocities of the Janus particles at each time step across far- and ``near-field regimes" (interparticle distances $\lesssim R$, but still at distances where slip layers do not overlap). 

\subsection{Fine-grained simulations}
\par Depending on the relative initial orientation of the particles $\Delta\theta = \theta_2 - \theta_1$ (Figs:~\ref{fig:ml_sch}b), we observe different types of trajectories. When two particles approach each other, they experience an effective attraction (Fig.~\ref{fig:sim_sch}b) either leading to binding or to scattering (Figs.~\ref{fig:ml_sch}b). 
For large interparticle distances (far field), the velocity of the particles decays approximately as $\sim 1/d$~(Fig. \ref{fig:sim_sch}b), where $ d = |\mathbf{r}_2 - \mathbf{r}_1| $, characteristic of phoretic interactions~\cite{nasouriExactPhoreticInteraction2020,SharifiMoodJFMPairInteraction2016,liebchen2019interactions}. In contrast, the near-field velocity exhibits a much steeper decay, scaling as $\sim 1/d^3$. Furthermore, at these closer distances, the relative velocity increasingly depends on the relative particle orientation
(Fig.~\ref{fig:sim_sch}b). Specifically, for larger $\Delta\theta$, the velocities are more widely scattered in the near field (see Fig. A1 in the Appendix).

\begin{figure*}
    \includegraphics[width=1.0\linewidth]{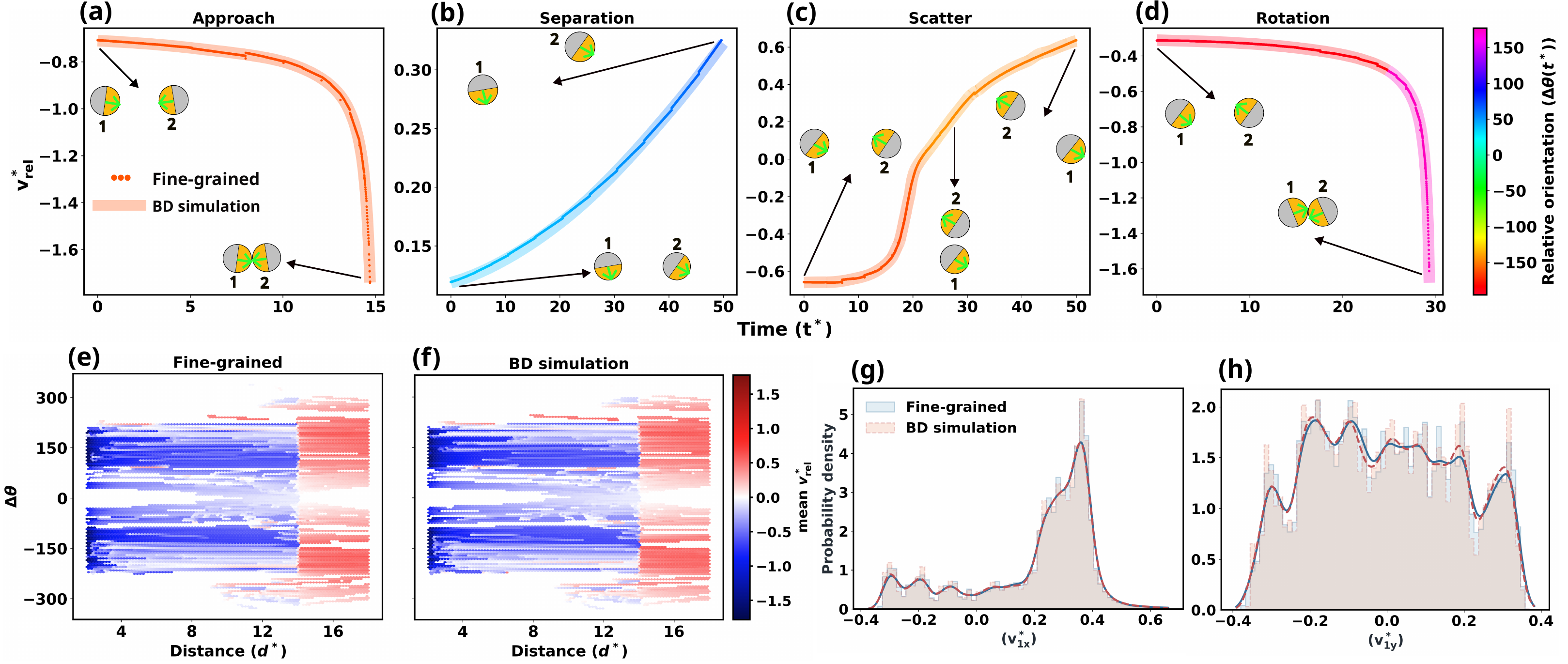}
    \caption{(a-d) Comparison of the temporal evolution of the relative velocity component $v^*_{rel}$ between full fine-grained and particle-based BD simulations for uniform (a-c) and nonuniform (d) surface mobility. Approach (a), where $v^*_{rel} < 0$, Separation (b), where $v^*_{rel} > 0$, Scatter (c), showing the transition from particle approach to separation, and Rotation (d). Trajectories are color-coded by the relative orientation ($\Delta\theta(t^*)$). Insets schematically illustrate particle orientations. (e-f) Histogram of relative velocity $v^*_{\mathrm{rel}}$ (color) as a function of inter-particle distance $d$ and relative orientation $\Delta\theta$ for fine-grained and coarse-grained BD simulations. Each bin is colored by the mean $v^*_{\mathrm{rel}}$ of all trajectory points falling within that bin. (g-h) Comparison of velocity distributions between fine-grained and coarse-grained BD simulations. Semi-transparent histograms show the normalized distributions, while the overlaid smooth curves show density plots for each dataset. Parameter for fine-grained model: $\mu=1$ (uniform), and $\mu=0$ (nonuniform). Parameters for coarse-grained model: $N = 2, \gamma=1, D_t = D_r = 0.0, \epsilon = 0.0$.}

    \label{fig:bd_two}
\end{figure*}

\subsection{Learning approach} 
 Following the previously outlined idea of this work, we now determine particle velocities and angular velocities directly from positions and orientations, without solving hydrodynamic and phoretic equations. To achieve this, we use a fully connected feedforward artificial deep neural network (DNN) (Fig.~\ref{fig:ml_sch}a). To construct a complete set of symmetry-invariant input features (feature vector), which we extract from the fine-grained simulations and feed into the input layer of the DNN, we first define the separation unit vector $\hat{\mathbf{r}} = (\mathbf{r}_2 - \mathbf{r}_1)/d$, its orthogonal complement $\hat{\mathbf{r}}_\perp$, and orientations vector \(\mathbf{p}_{1,2} = (\cos\theta_{1,2}, \sin\theta_{1,2})\). The input features consist of $d$, the relative angular components $\cos(\Delta\theta)$ and $\sin(\Delta\theta)$ (where $\Delta\theta = \theta_2 - \theta_1$), and the projections of the particle orientations onto the interparticle axis: $\hat{\mathbf{r}} \cdot \mathbf{p}_{1,2}$ and $\hat{\mathbf{r}}_\perp \cdot \mathbf{p}_{1,2}$. The angular terms encode the alignment (parallel vs. antiparallel) and the chiral offsets, respectively. These descriptors ensure invariance under global transformations while preserving all required information. Accordingly, the DNN predicts the angular velocities and the translational velocity components projected onto the relative displacement vector: $v_{i\parallel} = \mathbf{v}_i\cdot\hat{\mathbf{r}}$ and $v_{i\perp} = \mathbf{v}_i\cdot\hat{\mathbf{r}}_{\perp}$. This projection guarantees that the target outputs are frame-invariant. Crucially, to enable the construction of many-body dynamics via pairwise summation, we isolate the interaction effects by training on the `excess' velocity $\mathbf{v}_i - v_0 \mathbf{p}_i$, where $v_0$ is the self-propulsion speed determined from fully resolved single-particle simulations.
\par We use two independent, connected DNNs with identical architectures. The first network \(f_\Theta^{(V)}: \mathbf{x} \mapsto \mathbf{V}\) maps the feature vector \(\mathbf{x} \in \mathbf{R}^7\) to the velocities \(\mathbf{V} = (v_{1\parallel}, v_{1\perp}, v_{2\parallel}, v_{2\perp}) \in \mathbf{R}^4\), while the second network \(f_\Theta^{(\Omega)}: \mathbf{x} \mapsto \boldsymbol{\Omega}\) predicts the angular velocities \(\boldsymbol{\Omega} = (\Omega_1, \Omega_2) \in \mathbf{R}^2\).  
Each network consists of three hidden layers with $H_1 = 32, H_2 = 64, H_3 = 32$ nodes, and exponential linear unit (elu) activations~\cite{clevert2016fastaccuratedeepnetwork}. DNN parameters \(\Theta\) (weights and biases) are optimized separately by minimizing the mean squared error loss function, $\mathcal{L} = \frac{1}{N_\mathrm{batch}} \sum_{\mathrm{batch}} \|\mathbf{y}^\mathrm{pred} - \mathbf{y}^\mathrm{actual}\|^2,$ where $\mathbf{y}^{\mathrm{pred}} = f_\Theta(\mathbf{x})$ denotes the network output and 
$\mathbf{y}^{\mathrm{actual}}$ represents the ground-truth obtained from fine-grained simulations. Optimization is carried out using the Adam optimizer~\cite{kingma2017adammethodstochasticoptimization} with learning rate \(\eta = 10^{-3}\). Training is performed for $500$ epochs, and performance is evaluated on an independent test dataset to assess the network's ability to generalize to unseen data (see Figs: A2 a-f in the Appendix).
\begin{figure*}[ht]
	\includegraphics[width=1.0\linewidth]{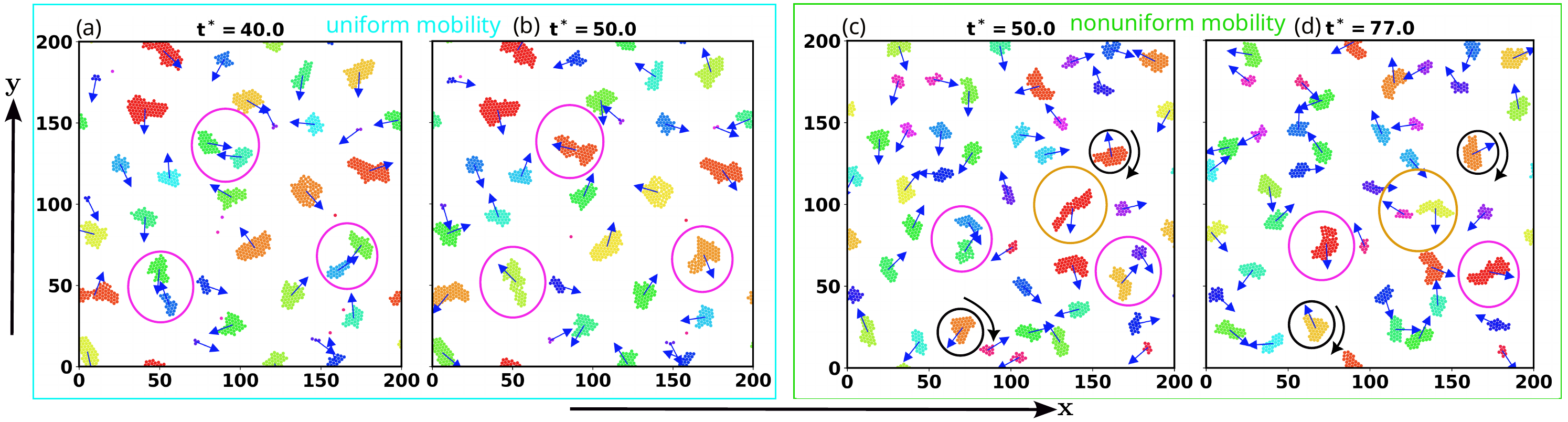}
	\caption{Snapshots from many-particle BD simulations with effective hydro-phoretic pair-interactions at packing fraction $\phi=0.1$ for uniform $(\mu=1)$ (a-b) and nonuniform $(\mu=0)$ (c-d) surface mobility. 
    Clusters are colored according to their IDs. Arrows indicate cluster translations and rotations. Over time, neighboring clusters merge (magenta circles), and for nonuniform surface mobility, they feature pronounced rotations (black circle), inducing their breakage and re-formation (orange circle). Parameters: $D_t = 0.01, D_r = 3D_t/4, \epsilon = 10$. The values of $D_t$ and $D_r$ are chosen such that $v_0/(R D_r)\sim 10^1-10^2$, as typical for autophoretic Janus colloids in water.} 
	\label{fig:bd_many}    
\end{figure*}
\subsection{Coarse-grained model}
To assess the accuracy of the presented coarse-graining method, we now numerically solve the following equations of motion (Active Brownian particle model), starting with the same initial configuration as in the fine-grained simulations.
\begin{eqnarray}
	\dot{\mathbf{r}}_i &=& v_0 \mathbf{p}(\theta_i) + \sum_{\substack{j=1 \\ j \neq i}}^{N}\mathbf{v}_{i,j}^{nn}  + \frac{1}{\gamma}\mathbf{F}_i^{\text{WCA}}+\sqrt{2D_t}\mathbf{\xi}_i\\
    \dot{\theta}_i &=& \sum_{\substack{j=1 \\ j \neq i}}^{N}\Omega^{nn}_{i,j} + \sqrt{2D_r}\mathbf{\zeta}_i
\end{eqnarray}
Here $\mathbf{v}_{i,j}^{nn} \equiv {v}^{nn}\left(|\mathbf{r}_i - \mathbf{r}_j|, \theta_i, \theta_j\right)$ and $\Omega_{i,j}^{nn} \equiv \Omega^{nn}\left(|\mathbf{r}_i - \mathbf{r}_j|, \theta_i, \theta_j\right)$ denote the linear and angular velocities predicted by the neural networks, respectively, $\gamma$ is the friction coefficient, and $N$ is the total number of particles. $\mathbf{F}_i^{\text{WCA}}$ represents the Weeks-Chandler-Anderson (WCA) force~\cite{WeeksJCPRoleRepulsive1971}, characterized by a repulsive force with strength $\epsilon=10$ and a cutoff distance $=2.07$, which is derived from the minimum inter-particle distance allowed in the two-particle fine-grained simulations.
$D_t$ and $D_r$ are the translational and rotational diffusion coefficients, respectively, while $\mathbf{\xi}_i$ and $\mathbf{\zeta}_i$ denote Gaussian white noise with zero mean and unit variance. We first analyze the component of relative velocity $v^*_{rel} = (\mathbf{v}_2 - \mathbf{v}_1)\cdot\hat{\mathbf{r}}$ for individual trajectories, plotted as a function of time. Depending on the sign of $v^*_{rel}$ and the surface mobility profile, we have identified four distinct types of dynamical scenarios: an approach phase, where $v^*_{rel} < 0$ signifies particles moving toward each other (Fig.~\ref{fig:bd_two}a); a separation phase, where $v^*_{rel} > 0$ indicates increasing separation (Fig.~\ref{fig:bd_two}b); a scattering phase, involving a transition in which particles initially approach one another ($v^*_{rel} < 0$), and subsequently move apart ($v^*_{rel} > 0$) (Fig.~\ref{fig:bd_two}c); and a rotation phase for nonuniform surface mobility, where the particles persistently rotate (Fig.~\ref{fig:bd_two}d). In Fig.~\ref{fig:bd_two}(a-d), these trajectories are color-coded by their relative orientation ($\Delta\theta(t^*)$), demonstrating that the coarse-grained model (thick, high-transparency solid lines) quantitatively reproduces both the spatial and the orientational dynamics of the fine-grained (scattered points) simulations. To achieve a more systematic overview and a first systematic comparison between fine-grained and coarse-grained simulations, in Figs.~\ref{fig:bd_two}e,f, we represent $v^*_{rel}$ (color) as a function of both distance and  $\Delta\theta$ for many different trajectories. Here again, the coarse-grained results are in close agreement with the fine-grained data. Finally, we compare the distribution of velocity components between the fine-grained and coarse-grained models, which are also in close quantitative agreement (Figs:~\ref{fig:bd_two}g,h and Figs: A3 a,b in the Appendix).

\subsection{Many particle Brownian dynamics simulations}
To demonstrate how our method can be used to explore the collective behavior of active colloids, within a few hours on a desktop computer, we now perform many-particle BD simulations $(D_t = 0.01, D_r = 3D_t/4, \epsilon = 10, \mu\in\{0,1\})$. We perform these simulations with random initial particle positions and orientations in a 2D box of size $L_x^* = 200$, $L_y^* = 200$ and periodic boundary conditions.
\par Perhaps unsurprisingly, for uniform mobility, $\mu=1$, we observe that the particles form dense clusters (Figs:~\ref{fig:bd_many}a,b and Video S1), even at a low packing fraction of $\phi = \frac{N\pi R^2}{4L_xL_y} = 0.1$. These clusters hardly rotate and slowly merge and grow (coarsen) over time. In contrast, for nonuniform mobility, $\mu=0$, we observe substantial cluster rotations 
~(Figs:~\ref{fig:bd_many}c,d, A4, and Video S2), which occasionally make them dynamically break and reform. This is in close analogy to the celebrated but still partly mysterious experiments showing living/dynamic clustering \cite{PalacciSLivingCrystals2013, GinotNCAggregationfragmentationIndividual2018,TheurkauffPRLDynamicClustering2012}.
This observation suggests that cluster rotations, induced by nonuniform surface mobility, could be key to finally understand the mechanism underlying dynamic clustering. Over long times, the individual cluster synchronize and all rotate clockwise or anticlockwise varying from simulation to simulation as will be studied separately. 
\section{Conclusions} 
The presented method, to learn effective interactions from fine-grained simulations, enables self-contained particle-only models of active particles with effective interactions representing hydro-phoretic pair interactions across near- and far-field regimes. 
These models could enable a new type of continuum theories and unprecedentedly large simulations of phoretic active matter. 
Notably, our method is highly generalizable 
and could help initiating a new era of formulating reduced models for complex nonequilibrium problems with complete timescale separation -- also beyond active matter.  
Generalizations in the near future 
could account for substrate-mediated osmotic interactions, many-body effects, several phoretic fields, and bulk reactions. 
Finally, to explicitly enforce momentum conservation in the reduced particle-only model, one could add 
constraints such as force-balance corrections or feedback mechanisms.

\begin{acknowledgments}
This project was funded by the Deutsche Forschungsgemeinschaft (DFG, German Research Foundation) in the framework of the collaborative research center Multiscale Simulation Methods for Soft-Matter Systems (TRR 146) under Project No. 233630050. 
\end{acknowledgments}
\bibliographystyle{apsrev_mod} 
\bibliography{ml_active}

\clearpage

\setcounter{equation}{0}
\setcounter{figure}{0}
\setcounter{table}{0}
\setcounter{page}{1}

\renewcommand{\theequation}{A\arabic{equation}}
\renewcommand{\thefigure}{A\arabic{figure}}

%
%
%
\section{APPENDIX }
\subsection{Details of fine-grained numerical solutions}
\par We numerically solved the fully coupled dimensionless equations of motion presented in the main text using the finite element software \textsc{COMSOL} Multiphysics with the PARDISO linear solver and an implicit backward differentiation formula (BDF) time-stepping scheme. A moving-mesh formulation accounts for the particle motion and shape deformation, while adaptive remeshing maintains numerical accuracy by refining the mesh whenever its quality falls below a predefined threshold~$0.1$. A very fine triangular mesh is used, with extra resolution in the interparticle region and near the particle boundaries.

\subsection {Distance-dependent velocity fit for the active colloids}
\par \noindent We fit the translational velocity of the Janus colloid, $v_1$, as a function of the interparticle separation $d = |\mathbf{r}_2 - \mathbf{r}_1|$ to the form
\begin{equation*}
	v_1^{*}(d^*) = \frac{a}{d^*{^n}} + v_0 ,
\end{equation*}
with $a$, $n$ and $v_0$ being the fit parameters and $d^* = d/R$ is the dimensionless separation distance. We fit the velocity only in the far-field regime, defined by $d > 5R$, where $R$ is the particle radius. In this regime, the fitted velocity profiles exhibit the expected $1/d$ scaling, consistent with long-range phoretic interactions (Fig.~2b, main text). At shorter separations, pronounced deviations from this scaling occur due to orientation-dependent near-field interactions. The fit parameters are obtained as follows: $a = 0.542 \pm 0.002,\ n = 0.969 \pm 0.004, \ \text{and} \ v_0 = 0.317 \pm 0.0003$ (far field); $a = 4.16 \pm 0.02,\ n = 3.02 \pm 0.005, \ \text{and} \ v_0 = 0.406 \pm 0.0002$ (near field). The value of $v_0$ in the far field matches closely with the self-propulsion speed of the colloids obtained from our fine-grained simulations of a single isolated colloid.
\begin{figure}[ht]
	\includegraphics[width=1.0\linewidth]{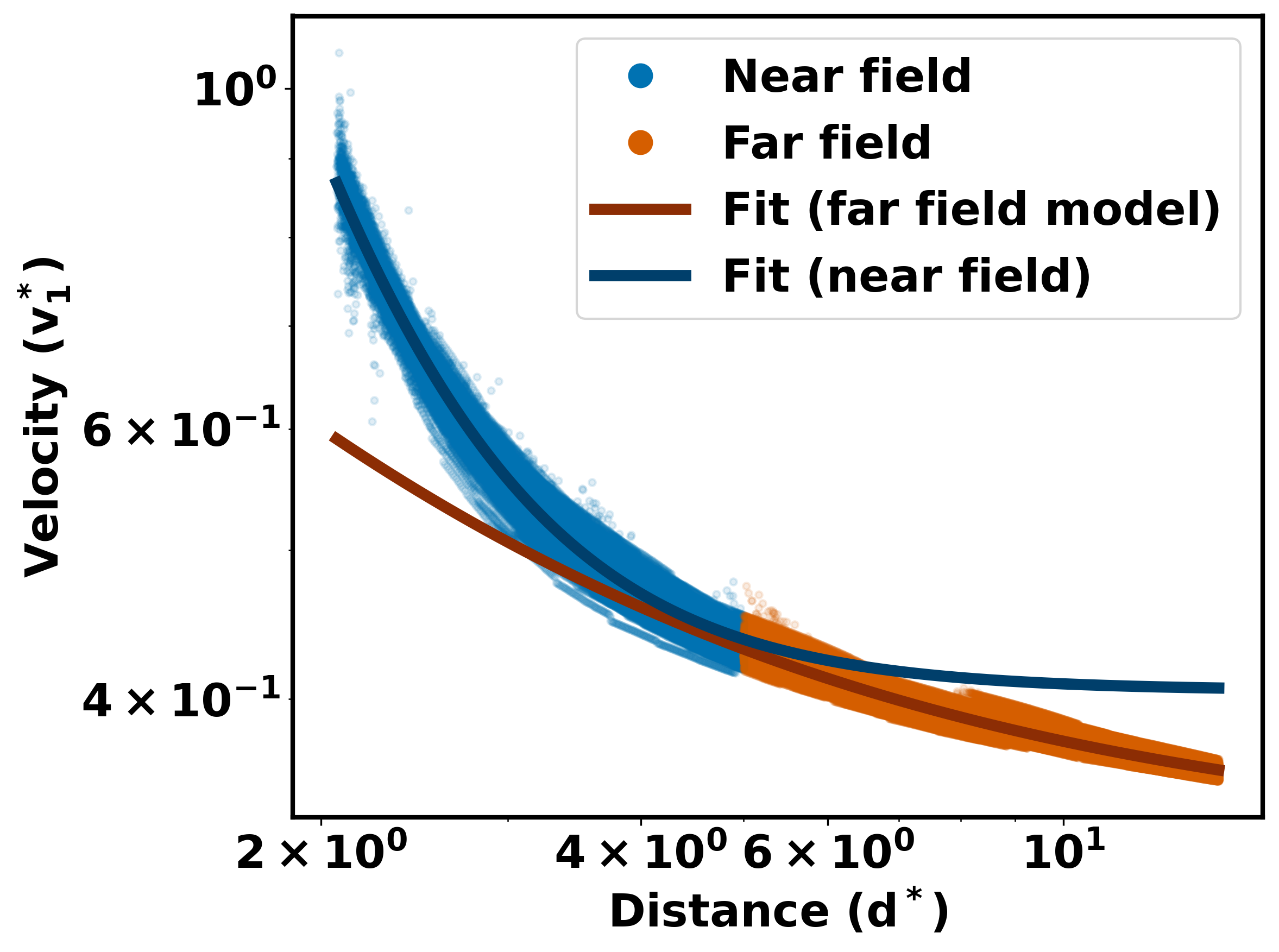}
	\caption{The fitted velocity profiles for $\Delta\theta \leq 20^0$ show a $1/d$ decay in the far field (dark orange line), whereas the near-field data reveal strong deviations from the scaling. Notably, the near-field velocity data exhibit significantly greater scatter than in the $\Delta\theta \leq 10^0$ case shown in the main text (Fig.~2b).}
	\label{fig:si_fit}
\end{figure}
\begin{figure*}[ht]
	\includegraphics[width=1.0\linewidth]{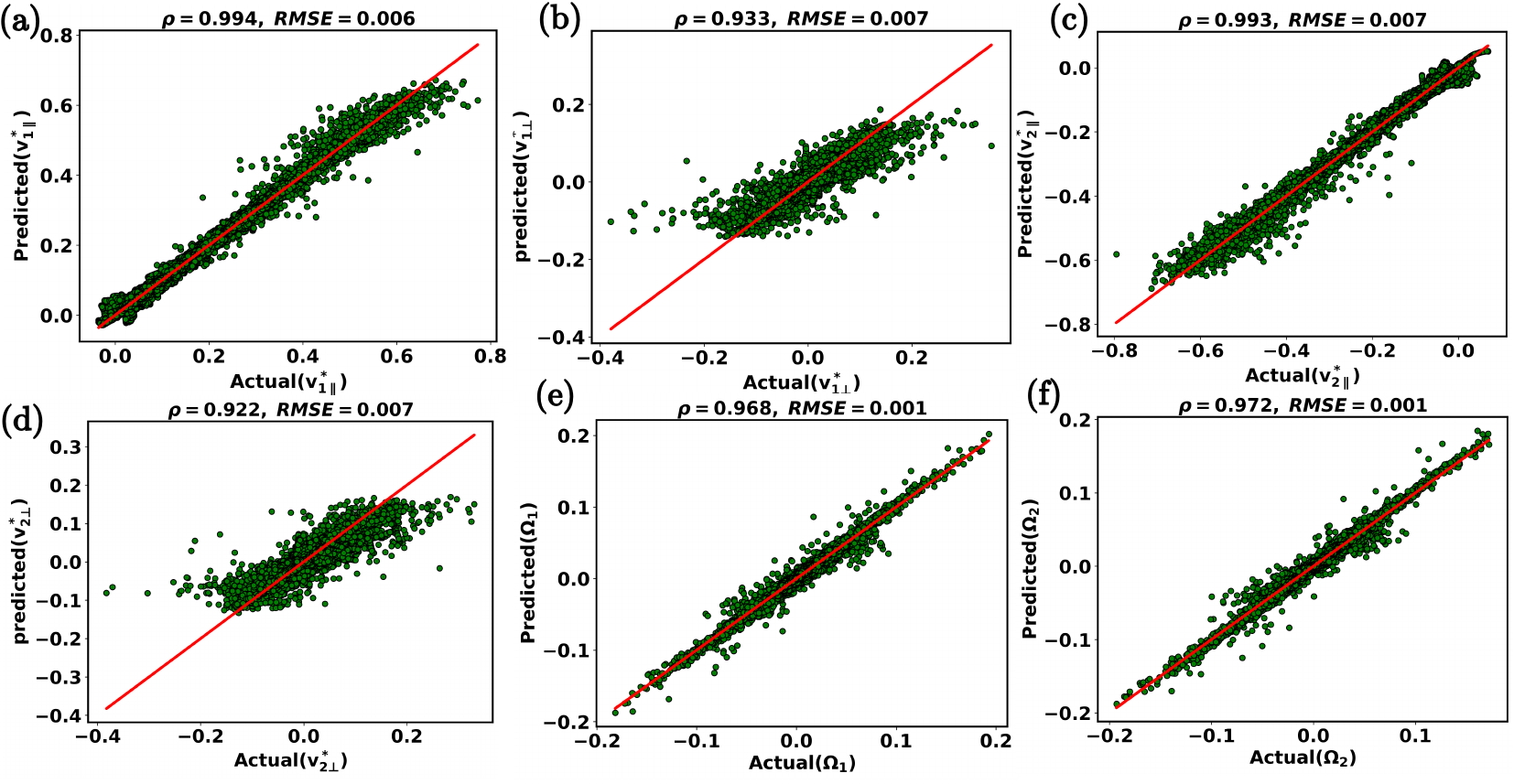}
	\caption{(a) Pearson correlation coefficient $\rho$ for the DNN’s predicted linear velocity (a-d) and angular velocity (e-f) components. The correlation between the true and predicted values for both linear and angular velocities is consistently close to unity across all test sets, indicating that the DNN effectively captures the underlying two-body hydro-phoretic interaction laws.}
	\label{fig:si_corr}
\end{figure*}
\subsection{Validation of DNN Predictions}
\par \noindent To quantify the predictive accuracy of the DNN, we compute the Pearson correlation coefficient
\(\rho = \frac{\mathrm{cov}(X^\mathrm{true}, X^\mathrm{DNN})}
{\sqrt{\mathrm{var}(X^\mathrm{true})\,\mathrm{var}(X^\mathrm{DNN})}}\), where \(X\) denotes either the velocity components of one of the particles \((v_{1\parallel},v_{1\perp},v_{2\parallel},v_{2\perp})\) or the angular velocities \((\Omega_1,\Omega_2)\). A value of \(\rho=1\) corresponds to perfect agreement, while \(\rho=0\) indicates random predictions. Additionally, we evaluate the root mean squared error, given by  \(\mathrm{RMSE} = \sqrt{\frac{1}{N} \sum_{i=1}^{N} (X_i^\mathrm{true} - X_i^\mathrm{DNN})^2}\), where \(N\) is the number of test samples, to quantify the deviation between the predicted and ``true" values from fine grained simulations. A lower RMSE value signifies better accuracy in the model’s predictions. As shown in Figs.~\ref{fig:si_corr}(a-f), \(\rho\) is consistently close to unity for both velocity components and angular velocities across all test sets, indicating that the DNN successfully captures the underlying two-body hydro-phoretic interaction laws. The RMSE is correspondingly low, confirming the model's strong predictive performance and its ability to generalize well to unseen data.

\subsection{Comparison between fine-grained and BD simulations}
\par
\begin{figure}[ht]
	\includegraphics[width=1.0\linewidth]{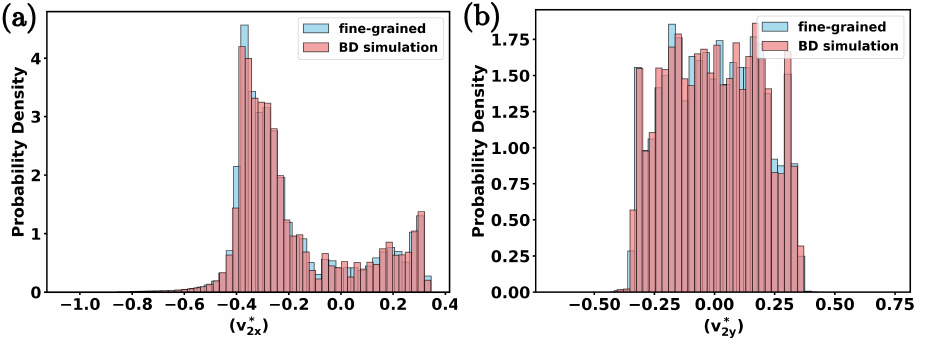}
	\caption{Comparison of velocity distributions between fine-grained and coarse-grained BD simulations for the 2nd particle. The velocity distributions for the active colloid in both simulations match closely, confirming that the DNN accurately captures the effective interactions without considering explicit field contributions.}
	\label{fig:si_vel}
\end{figure}

 We plot the probability distributions of the velocity components of the second particle obtained from fine-grained and coarse-grained Brownian dynamics simulations. Panels~\ref{fig:si_vel}(a,b) show the corresponding distributions for the two Cartesian velocity components for 2nd particle. The coarse-grained results are in close quantitative agreement with the fine-grained data, demonstrating that the coarse-grained model accurately reproduces the single-particle velocities.\\

\subsection{Spontaneous symmetry breaking}
\par
\begin{figure*}[ht]
	\includegraphics[width=1.0\linewidth]{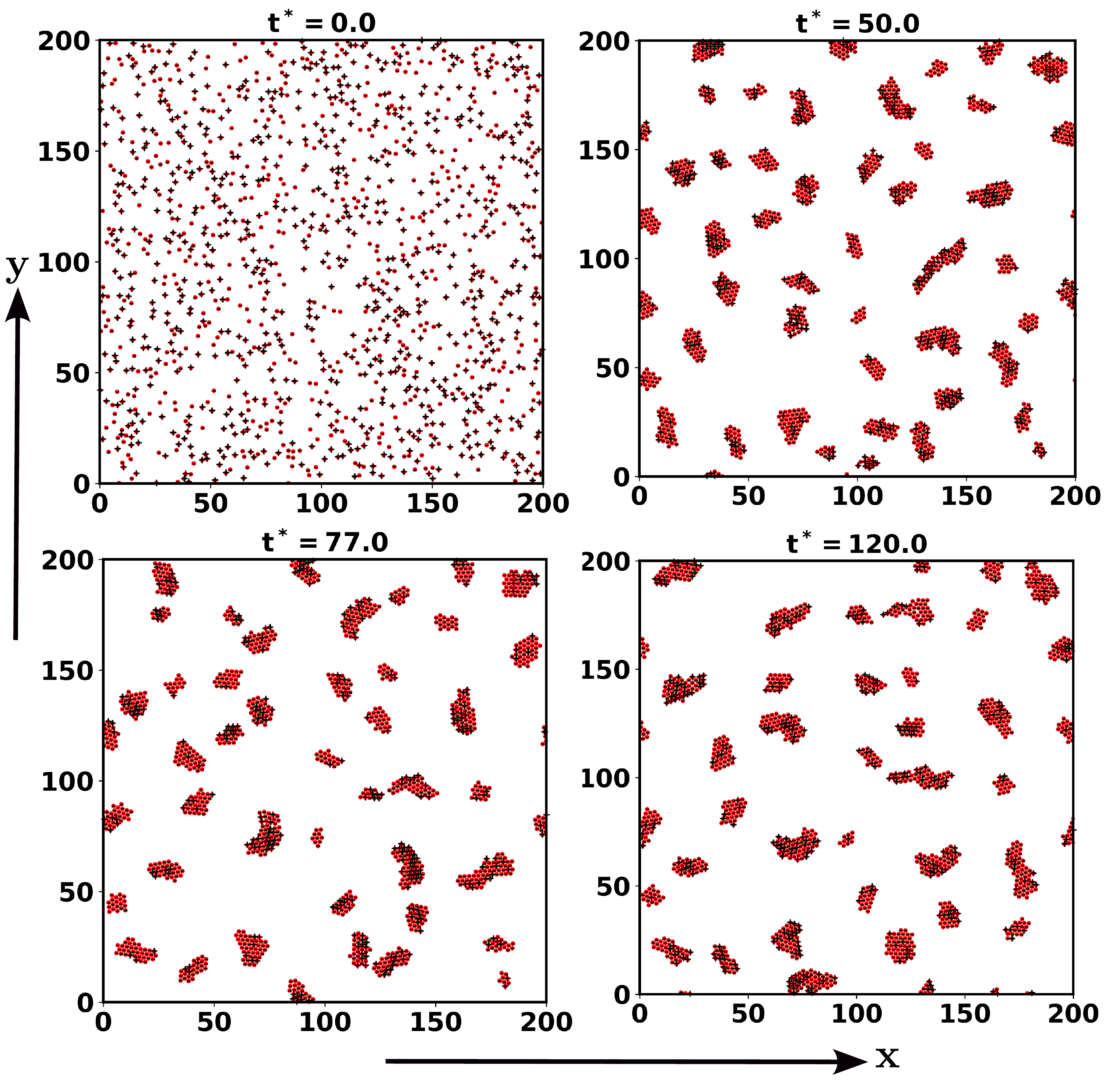}
	\caption{Emergence of rotating clusters due to spontaneous symmetry breaking for  nonuniform mobility. Time-lapse snapshots show the evolution from a disordered state with mixed angular velocities to clustered states with a dominant (clockwise) rotation, evidencing spontaneous rotational symmetry breaking.}
	\label{fig:si_many_snap}
\end{figure*}

 \noindent For nonuniform mobility, we observe the emergence of clockwise-rotating clusters, a phenomenon driven by spontaneous symmetry breaking. To confirm this, we analyzed time-lapse snapshots of the particles, tracking the evolution of the sign of their angular velocity ($\Omega$) (Fig.~\ref{fig:si_many_snap}). Initially, the system shows a disordered state with a balanced mixture of positive and negative $\Omega$ values. However, as the simulation progresses, the particles aggregate, causing a transition where some individual clusters adopt a predominantly negative sign. This shift demonstrates how the system breaks its initial rotational symmetry, leading to the formation of rotating clusters.
\subsection{List of movies}
\begin{enumerate}
	\item \textbf{Many body simulations of particles with uniform surface mobility ($\mu = 1$).}  
	Animated version of Fig.~4a,b (Video S1) showing the time evolution of many-particle Brownian dynamics simulations with effective hydro-phoretic interactions at packing fraction $\phi = 0.1$. The simulation parameters are  $D_t = 0.01$, $D_r = 3D_t/4$, and $\epsilon = 10$.
	\item \textbf{Many body simulations of particles with nonuniform surface mobility ($\mu = 0$).}  
	Animated version of Fig.~4c,d (Video S2) showing the corresponding many-particle Brownian dynamics simulations at $\phi = 0.1$. In this case, clusters exhibit pronounced rotations, occasionally leading to their dynamic breakage and re-formation. The simulation parameters are $D_t = 0.01$, $D_r = 3D_t/4$, and $\epsilon = 10$.
\end{enumerate}

\end{document}